\documentclass[twocolumn,showpacs,preprintnumbers,amsmath,amssymb,eqsecnum]{revtex4}

\usepackage{graphicx}% Include figure files
\usepackage{dcolumn}% Align table columns on decimal point
\usepackage{bm}% bold math

\begin{document}

\preprint{}

\title{Mott transition from a diluted exciton gas to a dense electron-hole
plasma in a single V-shaped quantum wire}

\author{T. Guillet}
 \altaffiliation[Also at ]{D\'epartement de Physique, Ecole Polytechnique
F\'ed\'erale de Lausanne, Switzerland}
 \email{guillet@gps.jussieu.fr}
\author{R. Grousson}
\author{V. Voliotis}
 \altaffiliation[Also at ]{Universit\'e Evry-Val d'Essonne, France}
\author{M. Menant}

\affiliation{Groupe de Physique des Solides, CNRS,\\
Universit\'es Pierre et Marie Curie et Denis Diderot, \\
 2 place Jussieu, F-75251 Paris Cedex 05, France }

\author{X.L. Wang}
 \altaffiliation[Also at ]{Department of Electrical Engineering, Yale
University, USA}
\author{M. Ogura}

\affiliation{Photonics Research Institute, National Institute of
Advanced Industrial Science and Technology (AIST), Tsukuba Central
2, Tsukuba 305-8568, Japan \\
 and CREST, Japan Science and Technology Corporation (JST), 4-1-8 Honcho,
Kawaguchi 332-0012, Japan}

\date{\today}% It is always \today, today,

\begin{abstract}
We report on the study of many-body interactions in a single high quality
V-shaped quantum wire by means of continuous and time-resolved
microphotoluminescence. The transition from a weakly interacting exciton
gas when the carrier density $n$ is less than $10^5 \ cm^{-1}$ (i.e. $n a_X
< 0.1$, with $a_X$ the exciton Bohr radius), to a dense electron-hole
plasma ($n > 10^6 \ cm^{-1}$, i.e. $n a_X > 1$) is systematically followed
in the system as the carrier density is increased. We show that this
transition occurs gradually~: the free carriers first coexist with excitons
for $n a_X > 0.1$, then the electron-hole plasma becomes degenerate at $n
a_X = 0.8$. We also show that the non-linear effects are strongly related
to the kind of disorder and localization properties in the structure
especially in the low density regime.
\end{abstract}

%Use showkeys class option if keyword
\maketitle

\section{Introduction}

Non-linear effects in one-dimensional (1D) systems have been recently the
subject of many theoretical as well as experimental investigations, due to
the potential applications for optoelectronic devices and especially for
incorporating low dimensional heterostructures, like quantum wires, in
lasers.

It has been predicted that due to the suppression of the 1D~singularity in
the density of states because of the important Coulomb correlations in one
dimension, the exciton transition has an enhanced oscillator strength and
stability compared to quantum wells \cite{Ogawa-PRB91, Rossi-PRB96}. Many
questions arise then: up to what carrier densities is the exciton stable~?
When does the Mott transition occur in a quantum wire (QWR) and what is the
origin of the emission line in 1D~systems when the carrier density is
increased~? There is a controversy in theory on this point, especially for
the critical carrier density at the Mott transition \cite{DasSarma-PRL00,
Piermarocchi-PRB01, Stopa-PRB01}. A typical predicted feature of a Mott
transition in optical experiments is the existence of gain in the
absorption spectrum at some energy range. But in some studies excitonic
lasing is told to be possible in one-dimensional structures. Rossi et al \cite{Rossi-PRB96} found optical gain at a carrier density of $4\ 10^6 \ cm^{-1}$
while the calculated Mott transition occurs at $8\ 10^5 \ cm^{-1}$. So, the
authors predict that lasing due to electron-hole pairs may occur in a QWR.
Piermarrochi et al \cite{Piermarocchi-PRB01} predict that optical gain
occurs at $10^5 \ cm^{-1}$, below the Mott transition, while Das Sarma et
al \cite{DasSarma-PRL00} found no gain at any carrier density. However in
photoluminescence experiments where the emission is observed, there is no
signature of the presence of gain in the structure, and it is more
difficult to evidence a Mott transition. The existence of the Mott
transition is still under debate, depending also on the definition itself.
In many-body theories treating the Coulomb correlations between free
carriers, the transition occurs when an electron and a hole cannot form a
bound pair and the obtained critical density is $n a_X \approx 0.1 - 1$
\cite{Rossi-PRB96, DasSarma-PRL00, Piermarocchi-PRB01, Stopa-PRB01} where
$n$ is the carrier density and $a_X$ is the exciton Bohr radius in the
system. However in many-body theories treating Coulomb interactions between
excitons, the transition occurs as soon as carriers (electron or hole) can
be exchanged between two excitons. Then the Mott criterion is very much
weakened, by a factor 100 in two-dimensionnal systems
\cite{Combescot-EPL01}.

Experimentally, optical non-linearities under high pump excitation have
been studied previously by other groups in less confining QWRs
\cite{Greus-EPL96, Cingolani-PRB93}. Band filling has been evidenced and
according to a theoretical modelisation of the results, band gap
renormalization has been observed. Ambigapathy et al
\cite{Ambigapathy-PRL97} have studied dynamics of excitons under high
density regime. The authors conclude that the band gap renormalization is
exactly compensating the exciton binding energy and the exciton remains the
stable excitation under very high pump power ($n = 3\ 10^6 \ cm^{-1}$, $n
a_X = 3$). The high density dynamics is also of crucial importance in the
understanding of the lasing mechanism in QWR structures \cite{Kim-PhysE00}.
Recently excitonic lasing in QWRs has been reported \cite{Wegsheider-PRL93,
Sirigu-PRB00, Rubio-SSCom01}, based on the observation that the emission
energy remains nearly constant within the inhomogeneously broadened
photoluminescence line. We believe that this experimental evidence is not a
sufficient criterion since at the high carrier density for which lasing is
obtained, an electron-hole plasma (EHP) is formed in the wire and emits at
the same energy. Therefore excitonic lasing in 1D~structures is still an
open question to our point of view.

We have performed systematic studies of continuous and time resolved
photoluminescence of a single quantum wire, as a function of carrier
density, by means of optical imaging spectroscopy. We show that the
different non-linearities observed, depend strongly on the kind of local
disorder in the wire especially in the low density limit. As pump power is
increased the exciton emission line gets broader but remains at a constant
energy position within a few $meV$. No distinct features appear in the
spectra, for instance a line that could be attributed to the EHP emission.
However we show that a gradual transition in the nature of excitations has
occured in the system. Indeed, the time-resolved photoluminescence
experiments in the high density regime reveal that the emission is due to
the formation of an EHP and not to excitons.

%The carrier density associated with this transition is evaluated to be
$1.2\ 10^6 \ cm^{-1}$, which is to our knowledge the first experimental
estimation of the Mott density in this system.

\section{Disorder effects in V-shaped QWRs}

The studied samples are $5 \ nm$ thick V-shaped
GaAs/Ga$_{0.57}$Al$_{0.43}$As QWRs grown by flow rate modulation
epitaxy on a $4 \ \mu m$ pitched V-grooved GaAs substrate
\cite{Wang-APL95}. These samples are very high quality structures
with strong confinement (the energy spacing between the first and
second subbands e1h1 and e2h2 is $60 \ meV$) and showing specific
1D~properties, like large optical anisotropy due to 1D~valence
band mixing. The localization of excitons in these QWRs has been
studied by microphotoluminescence ($\mu$-PL). The low temperature
$\mu$-PL confocal setup is composed of a microscope objective with
a large numerical aperture ($0.6$) which is fixed on a three axis
piezoelectric stage. This allows to scan images along a single
QWR. The spatial resolution is $0.8\ \mu m$, which is the limit of
diffraction of the microscope objective at 750 nm. The spectral
resolution is $40 \ \mu eV$ and the temporal resolution is $20 \
ps$. Excitation wavelength was always adjusted with a Ti-Sa or a
dye laser in order to create carriers only in the wire. This is
important for the correct calibration of the carrier density in
the wire. Details of the set-up can be found elsewhere
\cite{Bellessa-PRB98, Guillet-Loc}.

Our former studies have shown that at low temperatures, the optical
properties of excitons in  the previous generation of QWRs are governed by
localization effects leading to a discrete spectrum of states
\cite{Bellessa-PRB98}. Monolayer thickness fluctuations along the growth
axis occur, leading to the formation of local potential minima with a mean
size of $50 \ nm$ along the wire axis. Optical imaging spectroscopy has
allowed to characterize the local confining potential along the wire axis
and to relate the local disorder to the optical properties
\cite{Guillet-Loc}. Excitons are localized in such quantum boxes and at
very low pump power, the emission is composed of very sharp lines
corresponding to the emission of the lowest lying exciton states in each
box. The energy spectrum is discrete and the exciton dynamical properties
are those of a zero-dimensional (0D) system \cite{Bellessa-PRB98}. We say
then that the QWRs are in a 0D~regime. When the density of excitons created
in average is increased, different effects appear depending on the size of
the box along the wire axis. The density threshold for non-linear effects
being one exciton per box, the interaction between excitons will be weak if
they are trapped in different boxes and more effective when trapped in the
same box. Two different behaviours have been reported \cite{Wu-PRB00,
Bellessa-EPJB01}: biexciton formation or Auger scattering depending on the
size of the box. As the density is further increased, state filling effects
occur with an important broadening of the emission band due to
exciton-exciton or exciton-free carriers interactions.

\begin{figure}[t]
\begin{center}
\includegraphics{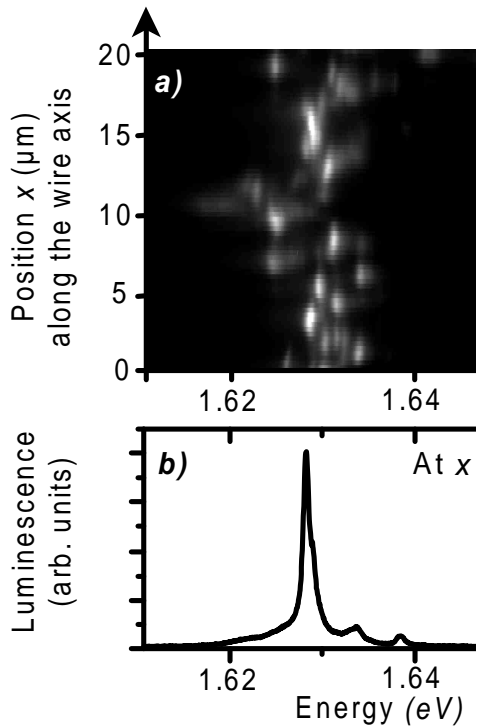}
\end{center}
\caption{a) Scanning optical image of a single QWR. The $\mu$-PL intensity
is represented in gray levels. b) Typical $\mu$-PL spectrum extracted from
the image at the position indicated by a white arrow.}
\label{fig:images}
\end{figure}

Recent progresses in the growth techniques allowed to further reduce the
heterointerfaces roughness of the QWR \cite{Wang-JCrystG00B}. In this new
generation of QWRs, the optical imaging spectroscopy shows that the
thickness fluctuations on the (001) facet occur every $5 \ \mu m$ while the
other heterointerfaces (311) and (111) present monolayer fluctuations from
$0.5$ to $2 \ \mu m$ \cite{Guillet-Loc}. A typical scanning image of a
single QWR is shown in figure~\ref{fig:images}. Each bright spot
corresponds to the emission from the lowest lying level in local potential
minima along the wire axis. The linear density of the emitting sites is
about 1 per $\mu m$ and the mean potential minimum length is $400 \ nm$ as
it is obtained after statistical analysis \cite{Guillet-Loc}. Sometimes the
emitting region can reach $3 \ \mu m$ on one wire, for example at the
position indicated by a white arrow on the image. In such a very long
island, the localization length is much larger than the thermal de Broglie
wavelength of the excitons, and the excitonic states form a quasi-continuum
of states leading to 1D~behaviour of the system. These samples are told to
be in a 1D~regime. The emission spectrum is then composed of a single
homogeneously broadened line as it shown on figure~\ref{fig:images}.
Another evidence for the 1D~character of the system is that the radiative
lifetime as a function of temperature shows the characteristic $\sqrt{T}$
dependence \cite{Guillet-Loc}. This behaviour reflects the $1/\sqrt{E}$
1D~local density of states.

%fig1

\section{Many-body effects in 1D~regime QWRs}

We have studied continuous and time-resolved µPL in different carrier
density regimes in the 1D~regime QWRs described above. Three density limits
can be distinguished if we compare the average distance between excitons
$1/n$, to the exciton Bohr radius $a_X$ which is equal to $70 \ \mbox{\AA}$
in our structure for a binding energy of $20 \ meV$ \cite{Guillet-X1D}.
When the density of excitons $n$ is very low then $n a_X <\!< 1$ and the
excitons form a weakly interacting gas, we say in that case to be in a
``dilute'' regime. On the opposite, the ``dense'' regime corresponds to
very high carrier density where $n a_X >\!> 1$ and in this case we expect a
dense EHP to be formed. Inbetween, there is also an ``intermediate'' regime
for which a Mott transition should occur in the system.

\begin{figure}[t]
\begin{center}
\includegraphics{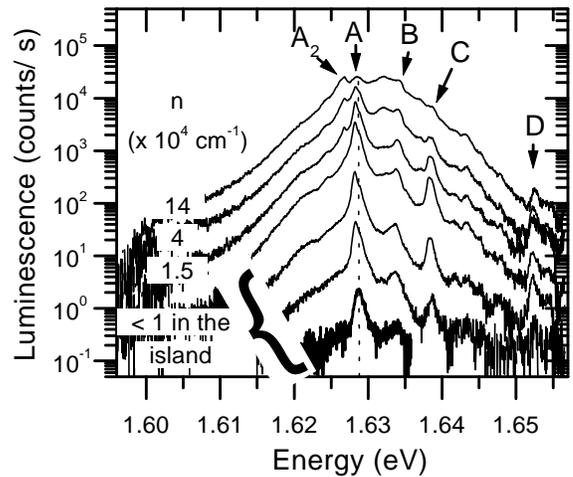}
\end{center}
\caption{$\mu$-PL spectra as a function of the pump power at $11 \ K$, under continuous excitation at $1.75 \ eV$ in the e3h3 transition of the wire. Peaks A and A$_2$ correspond respectively to the exciton and biexciton transitions of the considered island which is extended over $3 \ \mu m$, while B, C and D are associated to neighboring islands. The spectra correspond to excitation densities of 5400, 1700, 580, 180, 19, 2.2, 0.1 $W.cm^{-2}$ from top to bottom. The carrier densities are indicated for the 3 highest excitation powers.}
\label{fig:cw-spectra}
\end{figure}

Figure~\ref{fig:cw-spectra} represents typical $\mu$-PL spectra under
continuous excitation of a very long island (about $3 \ \mu m$) as a
function of carrier density in the low density regime. Excitation energy is
at $1.75 \ eV$ in the e3h3 transition of the QWR. At very low pump power
($P = 0.1 \ W.cm^{-2}$) one main $\mu$-PL peak appears, labelled~A on the
figure, corresponding to the emission of the lowest lying exciton level in
this island. As a matter of fact, two other peaks labelled~B, and~C appear
in the spectrum due to the presence of neighbouring islands that are
excited by the laser spot, as it has been shown by the scanning images of
the QWR (figure~\ref{fig:images}). However, their intensity is less than
peak~A intensity by a factor 10 and 30 respectively. As the pump power is
increased up to $5.4 \ 10^3 \ W.cm^{-2}$, there is a slight blue shift of
line~A by $0.2 \ meV$ and a new line appears in the spectrum corresponding
to the formation of a biexciton (labelled~A$_2$). The biexciton binding
energy can be deduced from the spectrum and is equal to $1.5 \ meV$ which
is in very good agreement with theoretical calculations \cite{Banyai-PRB87}
and similar to other experimental observations \cite{Crottini-PSSB00}. The
intensity of peak~A has a linear behaviour with pump power and saturates
above $10^4 \ W.cm^{-2}$ while peak~A$_2$ has a super-linear behaviour as
expected for biexciton. The total integrated intensity of the luminescence
line is linear with pump power. At the maximum pump power the lines get
broader and a broad background grows up. This background can be attributed
to the radiative recombination processes which are assisted by Coulomb
collisions between excitons or excitons and free carriers. Similar
observations have been reported by Vouilloz et al \cite{Vouilloz-SSCom98}.

%fig2

The calibration of the carrier density created in the QWR is made from the
data of saturation of exciton and biexciton emission intensities in the 0D
regime QWRs \cite{These}. We estimate that for a CW pump excitation at
$1.75 \ eV$, $400 \ W.cm^{-2}$ correspond to 1 exciton per~$\mu m$, i.e. a
carrier density of $10^4 \ cm^{-1}$ and to a number $n a_X$ equal to $7 \
10^{-3}$. This calibration has been used for the estimation of the carrier
density presented in all the following experiments.

As the upper limit of the dilute regime is reached ($n = 10^5 \ cm^{-1}$,
i.e. $n a_X \approx 0.1$) the lines associated to exciton and biexciton
transitions are still clearly marked showing that these are the stable
elementary excitations in the system.

\begin{figure}[t]
\begin{center}
\includegraphics{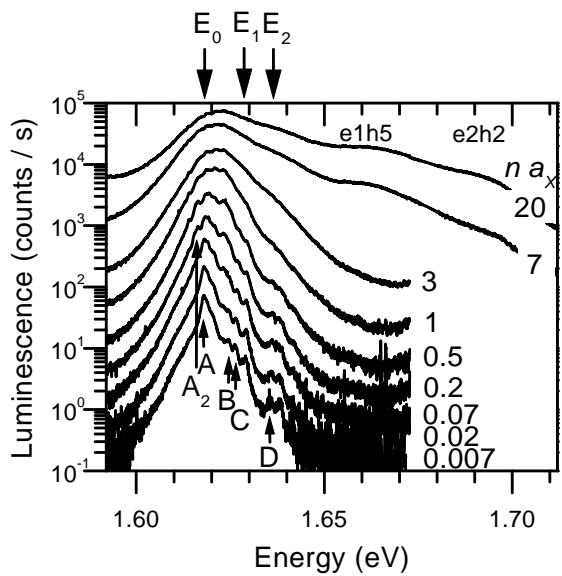}
\end{center}
\caption{$\mu$-PL as a function of carrier density under laser pulsed excitation at $11 \ K$, in an extended island similar to Figure~\ref{fig:cw-spectra}. The excitation energy is $1.77 \ eV$ in the e4h4 transition of the wire. The carrier densities $n a_X$ are indicated.}
\label{fig:ml-spectra}
\end{figure}

The intermediate ($0.1 < n a_X < 1$) and dense regime ($n a_X > 1$) are
represented in figure~\ref{fig:ml-spectra}. In order to study the dense
regime ($n a_X > 1$) pulsed laser excitation has been used to provide very
high carrier concentration in the wire up to $n \approx 10^7 \ cm^{-1}$,
i.e. $n a_X = 20$. For ($0.1 < n a_X < 1$) the spectra obtained under
continuous and pulsed excitation are similar. The peak labelled~A
corresponds to the main exciton line of a $2 \ \mu m$ long island. The
other peaks labelled~B, C, and D correspond as usual to neighbouring
islands excited by the tails of the laser spot. A$_2$ is the biexciton
emission line.

%fig3

Let us first focus on the intermediate regime. A qualitative change of the
spectra is observed above $n a_X = 0.1$. The exciton and biexciton lines
are broadened, they shift towards higher energies by $2 \ meV$ and finally
disappear. The shift of the lines is a signature of the many-body
interactions~: it corresponds to a competition between the band gap
renormalization (BGR) occuring as we fill the carriers subbands and the
screening of the exciton binding energy. The sum of both effects give rise
to small energy shifts in 1D~structures, only a few $meV$, moreover an
almost exact cancellation of these effects is predicted \cite{Rossi-PRB96,
DasSarma-PRL00}. The fast growing up of the broad background is attributed
to the presence of free electron-holes pairs. It progressively overcomes
the excitonic transition as the excitation power is increased.

\begin{figure}[t]
\begin{center}
\includegraphics{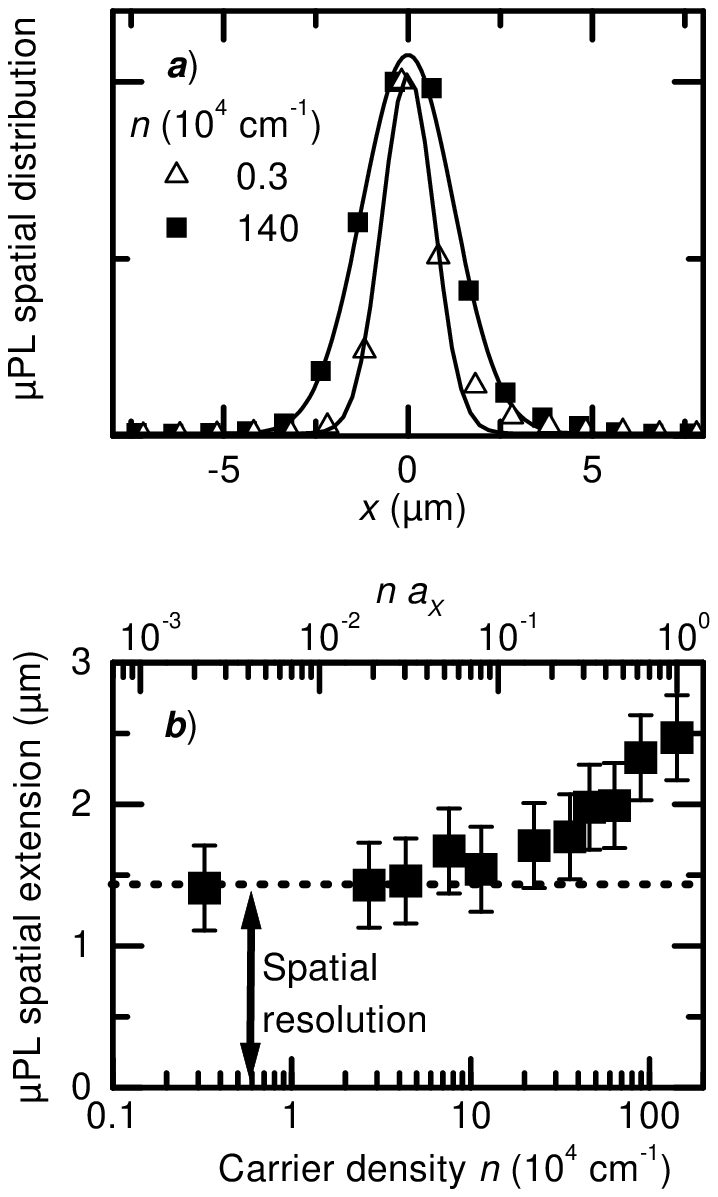}
\end{center}
\caption{a) Spatial profile of the luminescence emitted by an extended
island under continuous excitation
 at $x = 0$, for two carrier densities at $T = 11 \ K$. b) Width of the
fitting Gaussian curve is represented
  as a function of carrier density.}
\label{fig:diffusion}
\end{figure}

Another experimental evidence of the presence of free carriers in
the intermediate regime is that there is a spatial diffusion of
the carriers. Indeed we have measured the spatial distribution of
the emission as a function of the carrier density by the optical
imaging setup. In the case of 1D regime QWRs the emission spatial
distribution is about $1.2 \ \mu m$ thus larger than the limit of
resolution ($0.8 \ \mu m$) because of the larger spatial extension
of the islands ($400 \ nm$)\cite{Guillet-Loc}. The recorded
$\mu$-PL spatial profile can be very easily fitted by a Gaussian
curve whose width is reported in figure~\ref{fig:diffusion} as a
function of carrier density. For $n a_X < 0.1$, the spatial
distribution width is limited by our experimental resolution.
Above this density, the emitting region gets larger, showing that
diffusion of carriers takes place along the wire axis before
radiative recombination. We have checked that in the low density
regime, excitons do not diffuse even up to $70 \ K$ \cite{These},
because the exciton-exciton scattering mechanism is less efficient
than interaction with free carriers \cite{Honold-PRB89}. Here, the
observed diffusion for a carrier density above $10^5 \ cm^{-1}$ is
due to the presence of free carriers that are responsible for the
scattering with excitons. The temperature of the carriers deduced
from the high energy tail of the PL spectrum is $40 \ K$ at the
highest density, so that the diffusion isn't thermally activated
and is due to collisions between carriers. The density where this
process occurs is $n \approx 10^5 \ cm^{-1}$, i.e. $n a_X \approx
0.1$ and corresponds to the qualitative change observed in the
luminescence spectrum (figure~\ref{fig:ml-spectra})
\cite{Carrier-Density}. Both the diffusion of the carriers and the
spectral broadening of the exciton and biexciton transitions show
that free electrons and holes coexist with excitons and
progressively overcome them above $n a_X = 0.1$.

%fig4

The $\mu$-PL spectra reported on Figure~\ref{fig:ml-spectra} also present
the dense regime.
Above $n a_X = 1$, higher order subbands (e2, h2, h3, ...) are also
filled~: the optically allowed  transitions
e1h5 and e2h2 appear in the PL spectra at the same position as in the PLE
spectra. It is worth noticing that the
PL line centered at the exciton energy position does not shift
significantly up to the highest excitation power.
This is consistent with previous experimental findings
\cite{Cingolani-PRB93, Ambigapathy-PRL97} and reflects the
almost cancellation of BGR and screening of the exciton binding energy in
1D~systems.

\begin{figure}[t]
\begin{center}
\includegraphics{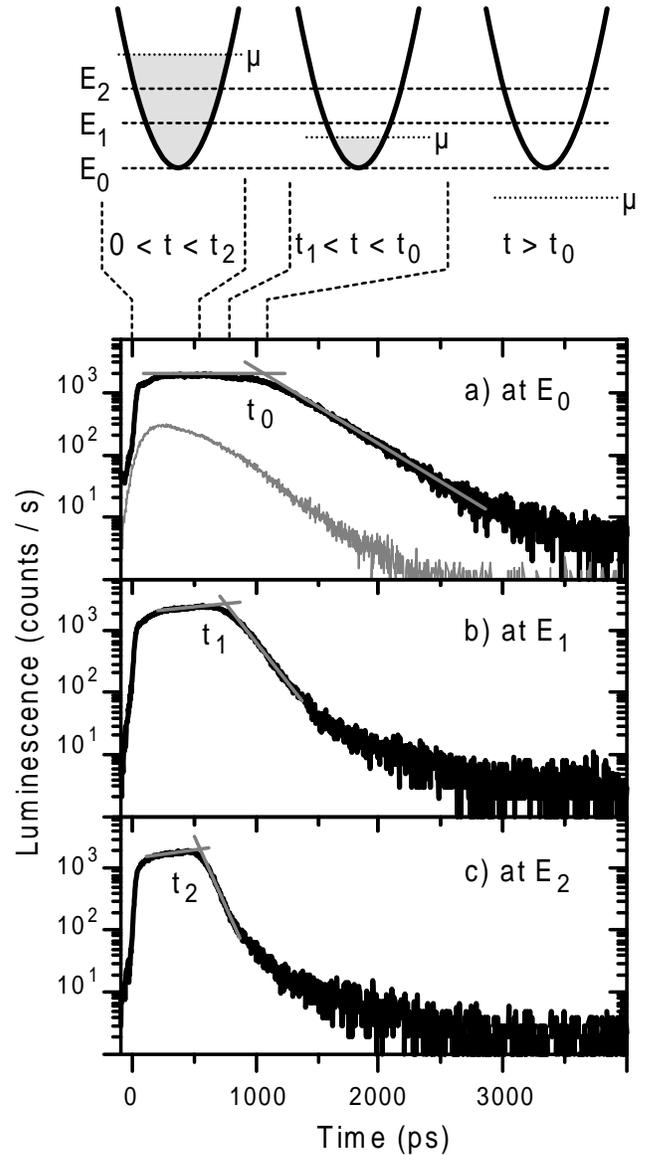}
\end{center}
\caption{Temporal evolution of the $\mu$-PL at high power excitation ($P =
8 \ 10^4 \ W.cm^{-2}$) detected at different energy positions in the line
($E_0=1.619 \ eV$, $E_1=1.629 \ eV$, $E_2=1.639 \ eV$) in the experiment
presented on figure~\ref{fig:ml-spectra}. The temporal evolution of the
chemical potential is schematically drawn on top in a simplified single
electron-hole band picture. The temporal evolution of the $\mu$-PL at $E_0$
at a lower pump power ($P = 8 \ 10^3 \ W.cm^{-2}$) is presented in gray
line in (a).}
\label{fig:trpl}
\end{figure}

The temporal evolution of the $\mu$-PL provides a more detailed
understanding of the system and is shown on
figure~\ref{fig:trpl} at the highest carrier density, corresponding to the
top curve of Fig.~\ref{fig:ml-spectra}
 ($n = 2.5 \ 10^7 \ cm^{-1}$, i.e. $n a_X = 20$, at $t = 0$). The
detection energy is set at different positions
 labelled~$E_0$, $E_1$, $E_2$ on the figure~\ref{fig:ml-spectra}. For the
three curves, we first observe a very
 fast rising time ($0 < t < 50 \ ps$), then a slowing down during $200 \
ps$ which corresponds probably to the
 setting up of the quasi-equilibrium in the system. Then a characteristic
saturation plateau is observed which
 persists longer as the detection energy is lower in the band.
 The PL intensity at the plateau is the same within 20\% at the three
detection energies, and a slight increase at
$E_1$ and $E_2$ is observed, which is not explained at the moment.
The temporal decay is different for the three curves. It is
non-exponential for the case of higher detection energies ($E_1$
and $E_2$) and reflects the complicated carrier dynamics at high
energy. When detecting on the band edge energy~$E_0$ after $1 \
ns$, the carrier density has decreased and we end up with an
exciton population that recombines. Thus the decay is nearly
mono-exponential with a characteristic time corresponding to the
exciton radiative lifetime (about $300 \ ps$)
\cite{Bellessa-PRB98} as also confirmed by the low density
temporal decay also shown on figure~\ref{fig:trpl}.

%fig5

The dependence of the saturation plateaux with detection energy is
characteristic of an EHP emission and not of exciton luminescence. Indeed
if this were the case, due to the conservation of the k-selection rule,
only $k = 0$ excitons could recombine and then the saturation should be
independent on the energy position in the line. On the contrary, we may
qualitatively interprete the PL temporal evolutions as follows~: at $t =
0$, as the laser pulse excites the system, the carrier density is about
$2.5 \ 10^7 \ cm^{-1}$ ($n a_X \approx 20$). An EHP out of equilibrium is
likely formed at this high density regime. Electron and hole subbands are
filled over hundreds of $meV$ up to the chemical potentials $\mu_{e,h}$,
and luminescence comes from the vertical transitions between electrons and
holes of the different subbands. In a simplified single electron-hole band
scheme described by only one potential $\mu$, the carrier density is
progressively reduced as recombination takes place, then $\mu$ decreases in
energy and the subband empties. When $\mu$ is larger than the detection
energy $E$ by an amount larger than $kT$, the occupancy factor at energy
$E$ is unity and the luminescence intensity saturates. When $\mu = E$, the
occupancy factor decreases and the luminescence decreases rapidly as well.
When $E - \mu >\!> kT$, the occupation factor at energy $E$ is a
Maxwell-Boltzmann distribution and the carriers at energy $E$ are at
thermal equilibrium with the band-edge. The carrier density corresponding
to $\mu = E_0$ is actually the density above which electron and hole
subbands are filled and the electron-hole plasma ends to be degenerate. The
whole process is schematically drawn on top of figure~\ref{fig:trpl}. The
time-resolved PL reflects in fact the chemical potential temporal evolution
and can be related to the carrier density. In a first order approximation
we may assume that the entire population decays with the same lifetime (the
exciton radiative lifetime). The carrier density we found in this way,
corresponding to $\mu = E_0$ is about $n \approx 1.2 \ 10^6 \ cm^{-1}$,
i.e. $n a_X \approx 0.8$. This threshold density corresponds to the
transition to a degenerate EHP, i.e. to the filling of the first electron
and hole subbands.

However in order to have a full description of the chemical potential
evolution as a function of $n$ in the degenerate EHP, one would have to
study the temporal decay of the whole PL line and know more precisely the
population decay. This would also allow to measure the precise shape of the
plateaux, which depends on the selection rules governing the recombination
process (conservation of energy, momentum, or both of them) and the
renormalization of electron and hole energy dispersion as a function of the
density.

In ref.~\cite{Ambigapathy-PRL97} where similar results have been obtained,
the authors present the time decay of the {\it integrated} PL in order to
analyse the radiative decay of the carriers. Their main conclusion is that
the recombination is dominated by ``excitonic correlations'' even at high
density ($3 \ 10^6 \ cm^{-1}$) since the decay is mono-exponential with the
characteristic exciton lifetime. We show here the energy- ad time-resolved
PL provides a more complete understanding of the dynamics and proves that a
fully degenerate EHP is formed at high density, excluding the presence of
excitons.

\section{Conclusion}

We have monitored the apparition of free carriers as well as the
degeneracy of the EHP in a QWR by an extended study of their spectroscopy
and dynamics, coupled to a local study relieving the inhomogeneous
broadening of the transitions. Our results clearly show that the energy
position of the emission line is not a sufficient criterion in order to
determine the nature of the excitations in the system, as it was previously
argued concerning the lasing mechanism in QWR lasers
\cite{Wegsheider-PRL93, Sirigu-PRB00, Rubio-SSCom01}. In the low density
regime the situation depends on the localization regime. In 0D~regime QWRs
with a localization length of the order of $50 \ nm$, a competition between
Auger effects and biexciton formation has been observed. But in 1D~regime
QWRs, in which each island is extended over $0.5$ to $3 \ \mu m$ and can be
considered as a small portion of really 1D~QWR, excitons are stable until
$n a_X \approx 1$, and they coexist with free charges above $n a_X \approx
0.1$. For $n a_X > 0.01$, biexcitons are formed and the slight shifts of
the exciton line are a signature of Coulomb interactions between excitons.
Above $n a_X =0.1$ ($ 10^5 \ cm^{-1}$) free electrons and holes coexist
with excitons, leading to a broad spectral background and the diffusion of
carriers along the wire axis. In the dense regime ($n a_X > 1$, $n > 10^6 \
cm^{-1}$) the saturation observed is the signature of a degenerate EHP
luminescence excluding the possibility of exciton formation and reflects
the temporal evolution of the chemical potential. Our results are in quite
good agreement with recent theoretical calculations which predict that no
new emission line attributed to the EHP should appear in the emission
spectrum. This is due to the fact that the energy shifts associated to the
screening of the exciton binding and to BGR occuring at high density are
small (few $meV$) in QWRs because of the compensation of these two effects.


\begin{thebibliography}{100}
\bibitem{Carrier-Density} The estimated carrier densities in the dense
regime have been corrected by the diffusion of carriers, especially in
fig.~\ref{fig:ml-spectra}.
\end{thebibliography}
\end{document}